# THE NEXT GENERATION OF FETs: CNTFETs

Roberto Marani and Anna Gina Perri

Electronic Devices Laboratory, Electrical and Information Engineering Department,
Polytechnic University of Bari, via E. Orabona 4, Bari – Italy
annagina.perri@poliba.it

*ABSTRACT*

*In this paper we present an exhaustive description* of *the basic types of CNTFETs. In particular we review two models, already proposed by us, which allow an easy implementation in circuit simulators, both in analog and in digital applications.*

## 1. INTRODUCTION

CNTFETs (Carbon Nanotube Field Effect Transistors) are novel devices that are expected to sustain the transistor scalability while increasing its performance. One of the major differences between CNTFETs and MOSFETs is that the channel of the devices is formed by Carbon NanoTubes (CNTs) instead of silicon, which enables a higher drive current density, due to the larger current carrier mobility in CNTs compared to bulk silicon [1]. In particular, with CNTs we obtain good operation even at very high frequencies [2-8].
A number of different geometries and solutions for CNTFETs, as well as their DC and low frequency behaviour [9-11], have been evaluated and reported in many papers.
In this paper we present an exhaustive description **of** the basic types of CNTFETs. In particular we review two models, already proposed by us [3-5], which allow an easy implementation in circuit simulators, both in analog and in digital applications.
The presentation of the paper is organized as follows. In Section 2 we briefly summarize the main characteristics of CNTs, while in Section 3 we examine the basic types of CNTFETs, analyzing, for each one, the principle of operation. In Section 4 a review of our two models is presented, while in Section 5 typical analogue circuits and logic blocks have been simulated both in Verilog-A and in SPICE, together with the discussion of relative results. The conclusions are described in Section 6.

## 2. REVIEW OF CNTs

A Carbon Nanotube, discovered in 1991 by S. Iijima, is a sheet of hexagonal arranged carbon atoms rolled up in a tube of a few nanometers in diameter, which can be many microns long. Graphene is a single sheet of carbon atoms arranged in the well known honeycomb structure [1] [12]. This lattice is shown in Fig. 1.
Carbon has four valence electrons, three of which are used for the $sp^2$ bonds. In $sp^2$-hybridization an electron is promoted from the 2s-orbital to a p-orbital, and then two electrons from different 2p-orbitals combine with the single electron left in the 2s-orbital to generate three equivalent $sp^2$-orbitals. These orbitals are planar with $120^o$ between the major lobes, and the remaining p-orbital is perpendicular to this plane. The leftover p-orbital is perpendicular to the graphene, and electrons in this orbital bond to other carbon atoms through weak π-bonds. The electrons in the p-orbitals are thus loosely bound and responsible for the conductance of graphite. Since the CNT is made up of one or more sheets of graphene rolled up in a tubular structure, the binding in the CNT is nearly identical to that of graphite. The differences in binding are due to the larger inter-shell distance in CNT compared to the interlayer distance of graphite, and the curvature of the graphene sheets [13-15].

Fig. 1 shows the construction of a graphene sheet, in which carbon atoms are located at each crossings and the lines indicate the chemical bonds, which are derived from $sp^2$-orbitals. $C_h$ is chiral vector, T is tube axis; φ is chiral angle [1].

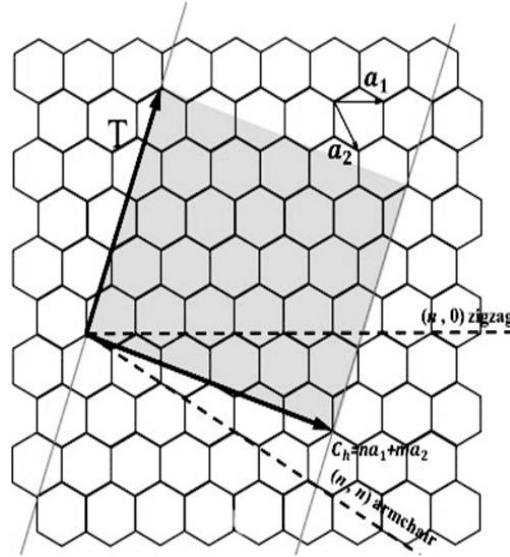

Figure 1. Lattice of graphene.

The chiral vector, $C_h$, is the vector perpendicular to tube axis T, which is given by:

$$C_h = n\bar{a}_1 + m\bar{a}_2 \tag{1}$$

being n and m a pair of integers and $\bar{a}_1$ and $\bar{a}_2$ the lattice vectors, which can be written as:

$$\bar{a}_1 = \left(\frac{\sqrt{3}}{2}a_0, \frac{3a_0}{2}\right) \qquad \bar{a}_2 = \left(-\frac{\sqrt{3}}{2}a_0, \frac{3a_0}{2}\right) \tag{2}$$

where $a_0$ is the iner-atomic distance between each carbon atom and its neighbor, equal to 1.42 Å.

The $p_z$ atomic-orbitals are oriented perpendicular to the plane and are rotational symmetric around the z-axis.

A CNT can be multi-wall (MWCNT) or single-wall (SWCNT) [1].

A MWCNT (Fig. 2) is composed of more than one cylinder whereas a SWCNT (Fig. 3) is a single cylinder.

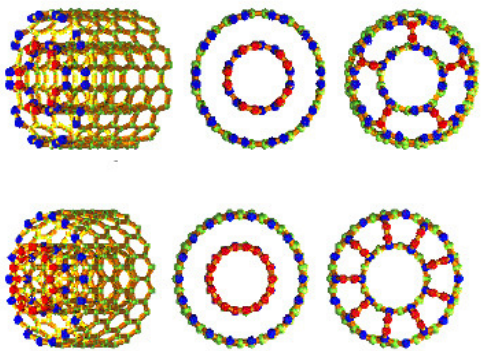 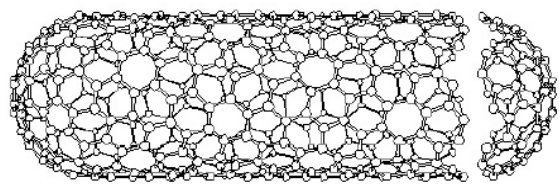

Figure 2. Structure of a MWCNT.     Figure 3. Structure of a SWCNT.

Depending on their chiral vector, CNTs have either semi-conducting or metallic behaviour.
In particular, if n = m or n – m = 3i, where i is an integer, the nanotube is metallic; in other cases it shows semi-conducting property [1] [12].
The diameter of the CNT can be calculated by the following equation [1]:

$$d = \left|\frac{C_h}{\pi}\right| = \frac{a_0}{\pi}\sqrt{n^2 + m^2 + nm} \qquad (3)$$

The chiral angle φ shows the chirality of nanotube and can be found by the following equation:

$$\cos\varphi = \frac{(n+m)\sqrt{3}}{2\sqrt{n^2 + m^2 + nm}} \qquad (4)$$

If (n = m, φ = 0°), CNTs are defined as armchair-type, while, if (m = 0, φ = 30°), as zig-zag type.
Fig. 4 shows the conduction band and valence band energy level diagram of carbon nanotube [16].

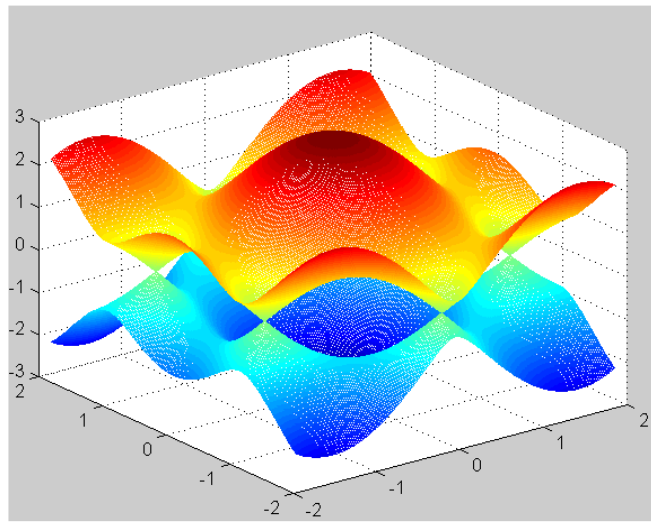

Figure 4. Energy level diagram of carbon nanotube.

In particular, for a SWCNT having a semiconductor behaviour, the band gap $E_{gap}$ is expressed by the following relation [1]:

$$E_{gap} = \frac{2\gamma_0 a_0}{d} \qquad (5)$$

where $\gamma_0$ is the carbon-to-carbon tight-binding overlap energy.
A straightforward application of this semiconducting property of CNTs is to form a field-effect transistor (FET) analogous to the MOSFET [1].

## 3. CNTFETs

The miniaturization has always a key role in electronic evolution: at each generation the miniaturization allows to obtain higher speed, lower power dissipation, lower costs and higher number of gates on chip.
Nowadays the known devices are very smaller, and a further reduction in size would give rise to tunnel effects thus degrading the whole performances. Therefore, the scientific community is looking for a new kind of devices, able to work better at nanometer scale, which is the ultimate limit in miniaturization.
Along with these new devices, molecular electronics will change the equation in our tool box, we will drop out well known partial differential equation for charge diffusion and we will use quantum mechanic to describe electrons, holes, atoms, molecules and photons. In coming years we will gain new tools from chemistry and physics, new sophisticated mathematical tool to include probability amplitude waves.

Carbon NanoTube Field Effect Transistors (CNTFET) are a new kind of molecular device. They are field effect transistors using a carbon nanotube as channel, and are regarded as an important contending device to replace conventional silicon transistors [17].

There are basically three types of CNTFETs [1]:
1. Schottky Barrier (SB) CNTFET
2. Partially-Gated (PG) CNTFET
3. MOS-like CNTFET (also known as C-CNTFET).

Early SB-CNTFETs have been typically p-type devices: the current carriers are holes and the devices are considered ON for negative gate bias. N-type CNTFETs can be obtained by direct doping of the tube with an electropositive element or by a simple annealing process of p-type CNTFETs.

To understand the operation of a **SB-CNTFET**, it is necessary to examine the energy band diagram for the structure. At the intersection between the metal contacts and the semiconducting carbon nanotube, Schottky barriers are created. The energy band diagrams are shown in Fig. 1 [1].

The current in CNTFETs is from the tunneling of carriers through the Schottky barriers. The type of metal for the contacts is chosen so that its work function forces the metal Fermi Level to lie between the valance and conduction band of the CNT [1].

For short channels, the CNT channel can become ballistic and hence, the metal contact resistance and the Schottky barriers at the source and drain ends limit the current drive through the nanotube. Thus, a low contact resistance, such as that of Titanium, is desirable. Presently, the control of the metal contacts to carbon nanotubes is not consistent and the tunneling current levels between transistors can vary greatly.

When a negative voltage is applied between the drain and source, the band structure, of the CNT, is modulated to account for the drain to source voltage ($V_{ds}$), as shown in Fig. 5.

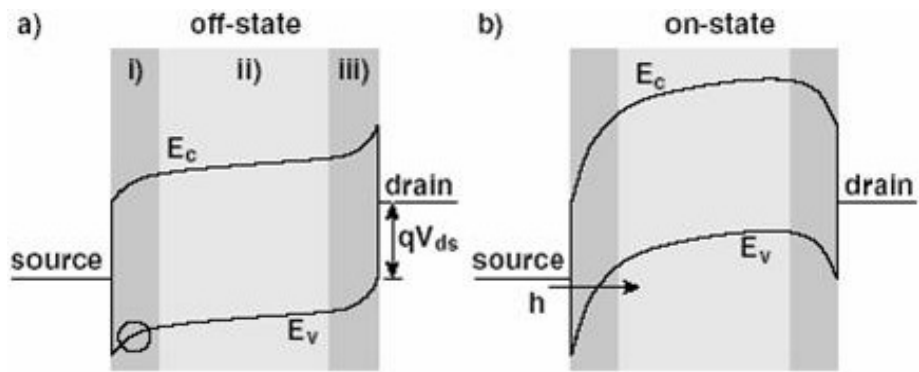

Figure 5. Band diagrams for a SB-CNTFETs in the 'off' and 'on' states respectively.

When a small negative gate to source voltage is applied, CNTFET is in the subthreshold condition.
With a negative gate voltage applied, the Schottky barrier width at the source is modulated, allowing for holes or electrons to tunnel through the valence band and pass to the drain. This condition is shown in Fig. 5b. The thickness of the source Schottky barrier at the metal Fermi level decreases exponentially with an increasing gate to source voltage. Thus, the tunnel current through the Schottky barrier increases exponentially, inversely to the barrier thickness.

In Fig. 6 the exponential current-voltage characteristics of the subthreshold condition is shown.
Moreover $I_d$-$V_{gs}$ characteristic does not differ greatly changing $V_{ds}$ because the drain voltage does not significantly control the source Schottky barrier.

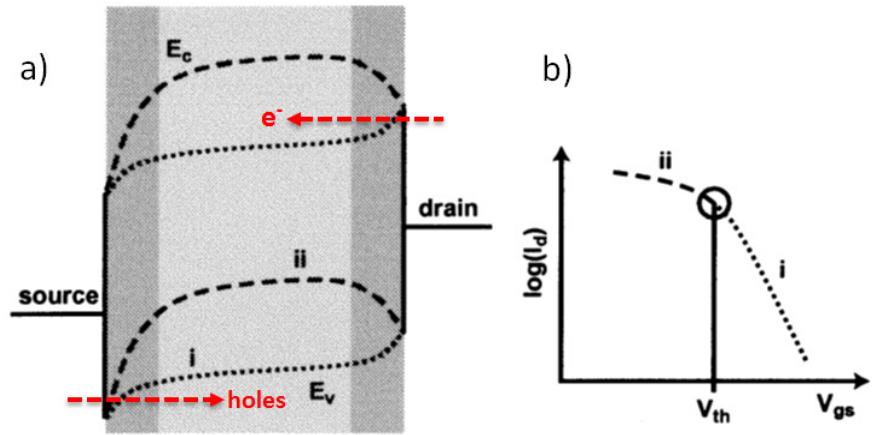

Figure 6. a) Subthreshold band diagrams for SB-transistors; b) I-V characteristics in subthreshold condition.

This exponential current relation can be seen in Fig. 7 [18], in which it is easy to see that the subthreshold characteristics do not vary largely with $V_{ds}$ and the subthreshold slope remains relatively constant with temperature.

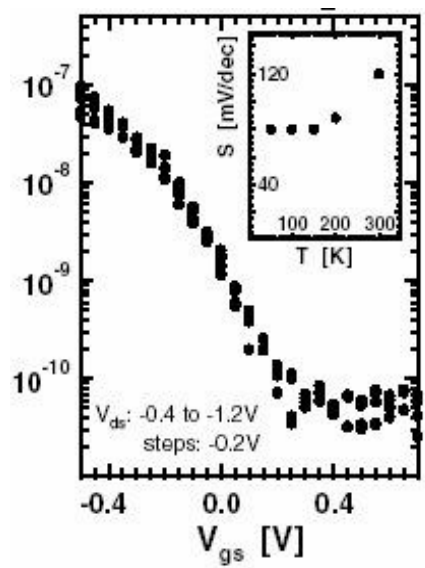

Figure 7. Experimental $I_d$ versus $V_{gs}$ subthreshold characteristics for a p-type transistor, with a channel length of 300 nm and a gate oxide thickness of 20 nm (*from [18]*).

The transistor threshold voltage, where the device acts similarly to an 'on' MOSFET, is reached when the metal source Fermi level is approximately even with the valence or conduction band of the CNT, in a p-channel or n-channel respectively. If the gate voltage continues to increase above this threshold, the Schottky barrier thickness at the source will remain constant and the current will not continue to increase exponentially. Above the threshold voltage, the current will only increase linearly with $V_{ds}$.
Above the CNTFET threshold voltage, the I-V characteristics are very similar to a MOSFET I-V characteristics: the current increases linearly with $V_{ds}$; and, when the barrier at the drain is completely eliminated, the FET current saturates.
The $I_d$-$V_{ds}$ characteristics for a saturated SB-CNTFETs have a very little slope, unlike short channel MOSFETs.
**PG-CNTFETs** are uniformly doped (or uniformly intrinsic) with ohmic contacts at their ends. PG-CNTFETs can be of n-type or ptype when respectively n-doped or p-doped.

These devices work in a depletion mode (uniformly n/p doped): the gate locally depletes the carriers in the nanotube and turns OFF the p-type device with an efficiently positive threshold voltage (efficiently negative for n-type) that approaches the theoretical limit for room-temperature operation. The ON current of such devices is limited by a "source exhaustion" phenomenon [19].

When the CNT is intrinsic, CNTFETs operate in enhancement mode and exhibit n- or p-type unipolar behaviour.

In case of **MOS-like CNTFETs** the source and drain are basically semiconductors p-type or n-type which are heavily doped [2-3]. These devices, also denoted as conventional CNTFETs or **C-CNTFETs**, show the best performances in terms of "on-off" ratio currents and subthreshold swing.

Fig. 8 shows a 3D representation of a C-CNTFET, whose conduction behaviour is similar to a common MOSFET.

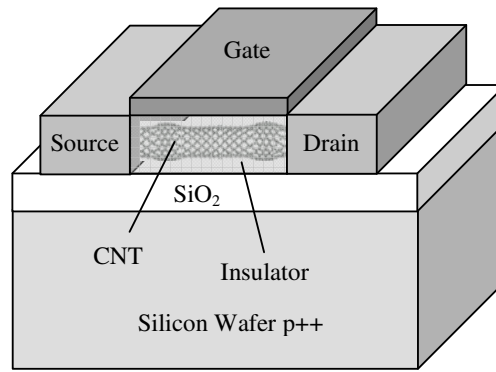

Figure 8. 3D representation of a C-CNTFET.

When a positive voltage is applied between drain and source ($V_{DS} > 0$ V), the hypothesis of ballistic transport allows to assert that the current is constant along the CNT and therefore can be calculated at the beginning of the channel, near the source, at the maximum of conduction band, where electrons from the source take up energy levels related to states with positive wave number, while the electrons from the drain take up energy levels related to states with negative wave number.

C-CNTFETs, with reference to SB-CNTFETs, show the following advantages:
1. unipolar characteristics and, therefore, faster;
2. reduction of the leakage current in the *off* state, due to absence of the Schottky barrier;
3. greater scalability;
4. a switch-*on* current of the source-channel junction significantly higher.

However, the control of the doping with ion implantation techniques are very difficult because the ions can replace the carbon atoms and destroy the desired properties of nanotubes.

## 4. MODELING OF CNTFETs

About modelling issues, most of the CNTFETs models available in literature are numerical and make use of self-consistency and therefore they cannot be directly implemented in modelling languages to design electronic circuits, such as SPICE, Verilog-A or VHDL-AMS. Moreover in the development of these new models we must face the problem of their simulation in real circuits.

As a general rule, the modelling of these new devices implies the solution of a set of partial differential equations. In this case the way to obtain correct results is to write a program written *ad hoc* for mathematical computation software (like Octave, Matlab), which allows in short time to obtain current-voltage characteristics of the simple device.

However, when we must simulate real complex circuits, we require the help of graphical interfaces to acquire circuit schemes and translated them in equations systems, which are typical tools of electronic simulations software. In particular, when the device behaviour can be expressed as a set of equations non involving PDE, using a compact model, it is possible to utilize some of the most useful tools available in electronic simulation software, having the component libraries and the graphic interface for the schematic drawing to obtain circuit netlist and circuit equations automatically. These both functions are the key point to reduce processing times, since, when circuits become complex, it is very difficult to solve and to check, one by one, a large system of integral-differential non linear coupled equations using, for example, Octave or Matlab. For this reason an electronic simulation software requires that the device should be described in an hardware description language.

The most widely used language available nowadays is SPICE, or any of its evolution, with various graphic interfaces. Another language having a solid implementation is Verilog-A, which is very interesting since it allows device description in a syntax quite near to C programming language.

In this section we review a compact, semi-empirical model, already proposed by us [3] [6-7], in which we have introduced some improvements to allow an easy implementation both in SPICE and Verilog-A simulator.

We also want to emphasize that a device model is considered a *compact* model because of the methods used to develop the equations and coefficients used for the electrical representation of the physical behaviour of a device. The word *compact* is used because these equations are simplified based upon several assumptions that are made when developing the model equations. On the other hand the availability of accurate, robust, and efficient compact models is fundamental to the successful utilization of any circuit simulation tool.

**4.a I-V model**

An exhaustive description of our model is in [3] [6-7]. In this Section we just describe the main equations on which is based our model.

When a positive voltage is applied between drain-source ($V_{ds} > 0$ V), the previous hypothesis allows to assert that the current is constant along the CNT and therefore it can be calculated at the beginning of the channel, near the source, at the maximum of conduction band, where electrons from the source take up energy levels related to states with positive wave number, while the electrons from the drain take up energy levels related to states with negative wave number.

When a positive voltage is applied between gate and source ($V_{gs} > 0$ V), the conduction band at the channel beginning decreases by $qV_{CNT}$, where $V_{CNT}$ is the surface potential and q is the electron charge. With the hypothesis that each sub-band decreases by the same quantity along the whole channel length, the drain current for every single sub-band can be calculated using the Landauer formula [20]:

$$I_{dsp} = \frac{4qkT}{h}\left[\ln(1+\exp\xi_{Sp}) - \ln(1+\exp\xi_{Dp})\right] \qquad (6)$$

where k is the Boltzmann constant, T is the absolute temperature, h is the Planck constant, p is the number of sub-bands, $\xi_{Sp}$ and $\xi_{Dp}$ have the following expressions:

$$\xi_{Sp} = \frac{qV_{CNT} - E_{Cp}}{kT} \quad \text{and} \quad \xi_{Dp} = \frac{qV_{CNT} - E_{Cp} - qV_{DS}}{kT}$$

being $E_{Cp}$ the sub-bands conduction minima.

Therefore the total drain current can be expressed as [20]:

$$I_{ds} = \frac{4qkT}{h}\sum_{p}\left[\ln(1+\exp\xi_{Sp}) - \ln(1+\exp\xi_{Dp})\right] \qquad (7)$$

To evaluate the surface potential $V_{CNT}$, in [3] the following approximation has been proposed:

$$V_{CNT} = \begin{cases} V_{gs} & \text{for } V_{gs} < \frac{E_C}{q} \\ V_{gs} - \alpha\left(V_{gs} - \frac{E_C}{q}\right) & \text{for } V_{gs} \geq \frac{E_C}{q} \end{cases} \qquad (8)$$

where $E_C$ is the conduction band minimum for the first sub-band and $V_{gs}$ the gate-source voltage. The parameter α, depending on $V_{ds}$ voltage, CNTFET diameter and gate oxide capacitance $C_{ox}$, has been extracted from the experimental device characteristics [2-3].

Fig. 9 compares the $I_{ds} - V_{ds}$ characteristics (denoted by continuous lines) of numerical simulations with Verilog-A language [21] and the experimental ones [22] (denoted by ●).

We have implemented the static characteristics both in SPICE and in Verilog-A, obtaining practically the same trends [2-3].

Moreover the proposed analytical modelling equations describing the current transport in CNTFETs have been developed from physical electronics [23-24].

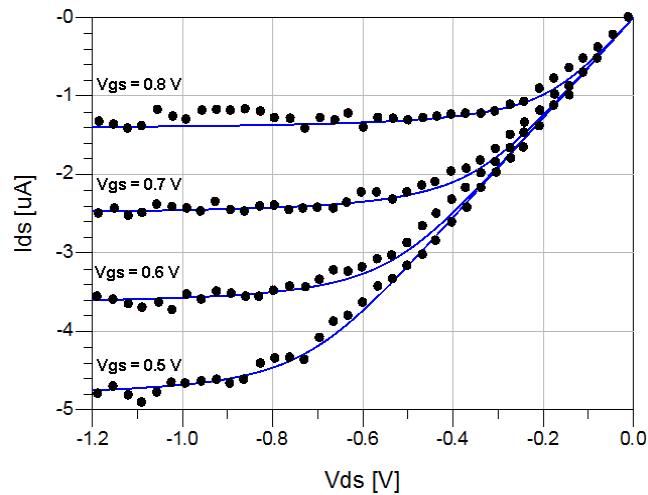

Figure 9. Simulated $I_{ds} - V_{ds}$ characteristics (denoted by continuous lines) and experimental ones [22] (denoted by ●) (*from [2]*).

### 4.b C-V model

For the dynamic analysis, it is necessary to determine the quantum capacitances $C_{GS}$ and $C_{GD}$, and therefore to know the total channel charge $Q_{CNT}$, which has the following expression:

$$Q_{CNT} = q \sum_p (n_{Sp} + n_{Dp}) \qquad (4)$$

where $n_{Sp}$ and $n_{Dp}$ are electron concentrations by the source and the drain respectively in the p-th sub-band. Omitting all the mathematical passages, exhaustively described in [2-3], the quantum capacitances $C_{GD}$ and $C_{GS}$ are given by:

$$\begin{cases} C_{GD} = q \sum_p \dfrac{\partial n_{Dp}}{\partial V_{gs}} = q \sum_p \dfrac{\partial n_{Dp}}{\partial \xi_{Dp}} \dfrac{\partial \xi_{Dp}}{\partial V_{CNT}} \dfrac{\partial V_{CNT}}{\partial V_{gs}} \\ C_{GS} = q \sum_p \dfrac{\partial n_{Sp}}{\partial V_{gs}} = q \sum_p \dfrac{\partial n_{Sp}}{\partial \xi_{Sp}} \dfrac{\partial \xi_{Sp}}{\partial V_{CNT}} \dfrac{\partial V_{CNT}}{\partial V_{gs}} \end{cases} \qquad (5)$$

The CNTFET equivalent circuit, reported in Fig. 10, is characterized by the generator $V_{FB}$, accounting the flat band voltage, the resistors $R_D$ and $R_S$, which include the parasitic effect due to the electrodes, the quantum capacitances, computed from the charge in the channel, and the CNT quantum inductance, assumed constant (equal to 4 pH/nm).

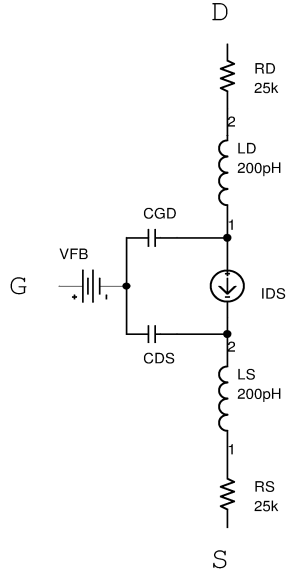

Figure 10. Equivalent circuit of a n-type C-CNTFET (*from [3]*).

**4.c Subthreshold Characteristics modeling of CNTFETs**

With reference to a SB-CNTFET, the total current in a CNTFET, as described previously, depends on the tunnelling current both of the holes and of the electrons through the source and drain Schottky barriers: the current in a SB-CNTFET exponentially increases, reducing the thickness of the Schottky barrier at source, with also an increase of the subthreshold current.

In comparison with MOSFETs, in CNTFETs we have not the minimum current for $V_{gs} = 0$ V, but to $V_{gs} = V_{ds}/2$. This is true for all SB-CNTFETs having the same metal used for the gate, drain and source. This minimum condition ($V_{gs} = V_{ds}/2$) is independent on metal work function.

In [5], to model the $I_{ds}$-$V_{gs}$ characteristics, we have implemented the equations proposed in [5] in a Matlab code, characterized by a minimum current at $V_{ds}/2$, threshold voltage and exponential subthreshold current.

In this paper we only report the simulated $I_{ds}$-$V_{gs}$ characteristics, for a n-channel and a p-channel CNTFETs, shown respectively in Fig. 11 and in Fig. 12.

For an exhaustive description of the main equations on which is based the model of CNTFET operating in subthreshold regime, we refer the reader to examine Reference [5].

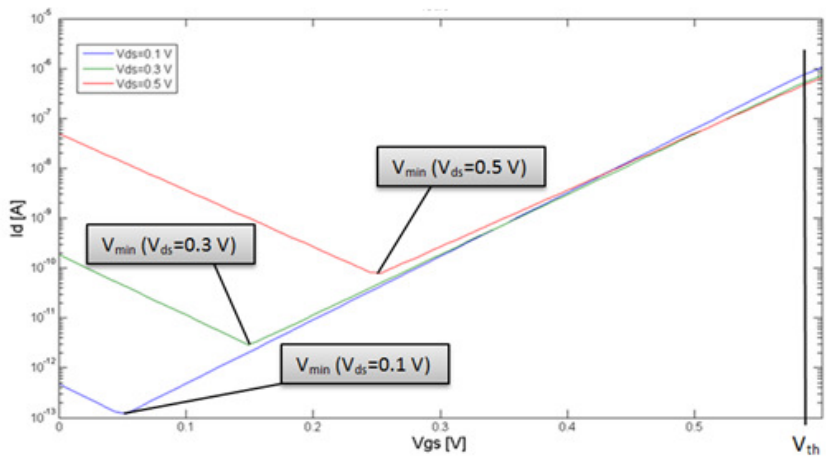

Figure 11. Simulated $I_{ds}$-$V_{gs}$ characteristics of a n-channel SB-CNTFET.

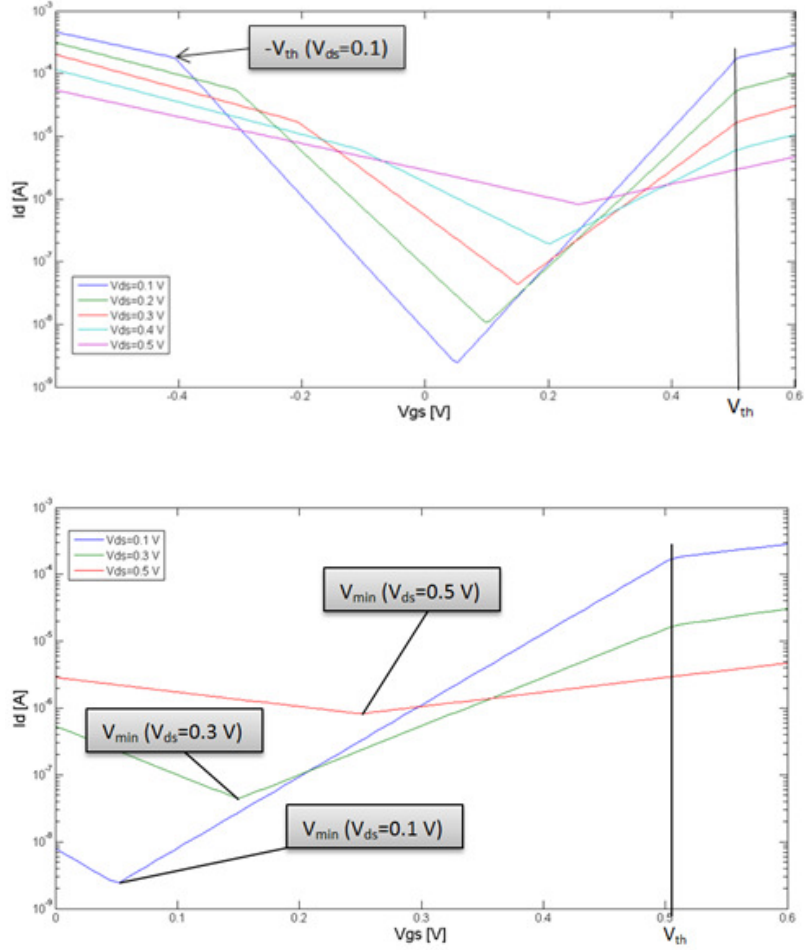

Figure 12. Simulated $I_{ds}$-$V_{gs}$ characteristics of a p-channel SB-CNTFET.

With reference to Figs, 11 and 12, the CNTFET subthreshold regime is verified between the two threshold voltage ($V_{th}$) points. These points have the same distance from the minimum voltage.
The distance from the minimum to the threshold is $\Delta V_{SUB}(V_{ds}) = V_{th} - V_{min}$.
The proposed procedure [5] allows to model the current in SB-CNTFETs., operating in the low voltage, subthreshold condition, and allows to determine the functionality of future subthreshold CNTFETs in digital design.

## 5. DESIGN OF TYPICAL ANALOG AND LOGIC CIRCUITS

In order to verify the versatility of our models, previously described, we have employed it to design both analog and digital circuits.
In all simulations we have considered CNTFETs having 1.42 nm of diameter, 100 nm length and quantum capacitances depending on polarization voltages.
The first example is a CNTFET amplifier in common-source configuration, shown in Fig. 13, in which we have reported circuit data are reported.

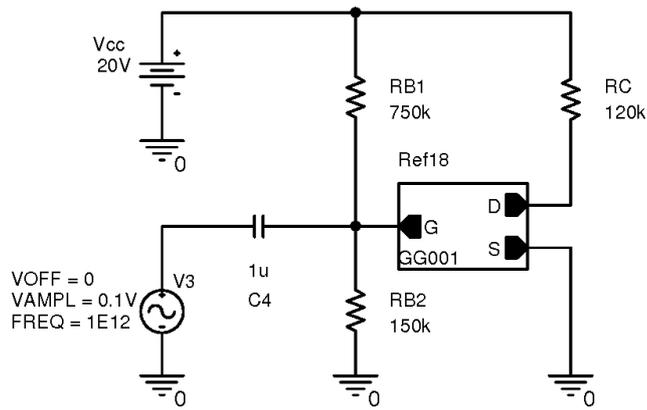

Figure 13. A CNTFET amplifier in CS configuration.

In Fig. 14 we have shown the Bode diagrams of the simulated amplifier with SPICE, while in Fig. 14 the same diagrams obtained with Verilog-A. In both cases the resistances of the doped drain and source regions are assumed equal to 25 kΩ.

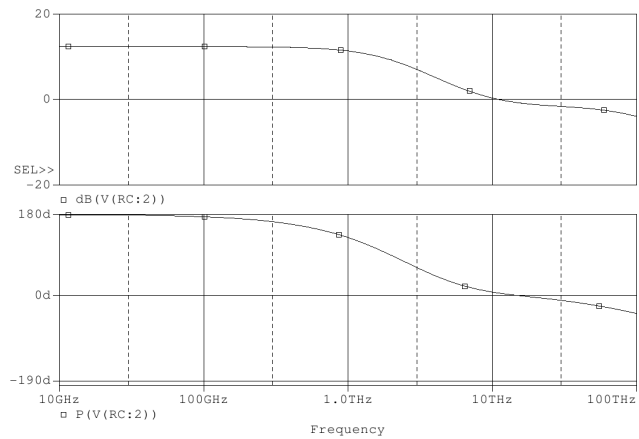

Figure 14. Bode diagrams of CS CNTFET amplifier with SPICE.

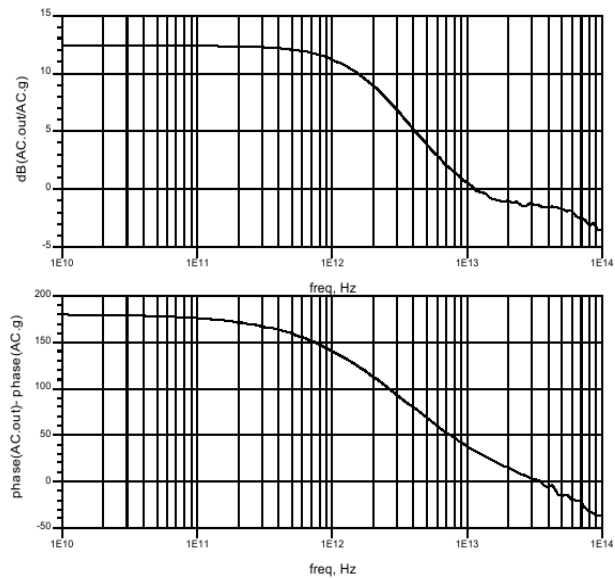

Figure 15. Bode diagrams of CS CNTFET amplifier with Verilog-A .

It is easy to see that this circuit shows a quite good agreement between the two model implementations, with small differences at frequency higher than the cut-off frequency.

It is important to underline that we have obtained the previous two figures considering in the CNTFET model also the classical inductance of 400 pH, which we have splitted up into two inductances of 200 pH in the source and drain terminals respectively, as shown in Fig. 10.

As second example, we have studied NAND gates in a five stage chain, as shown in Fig. 16

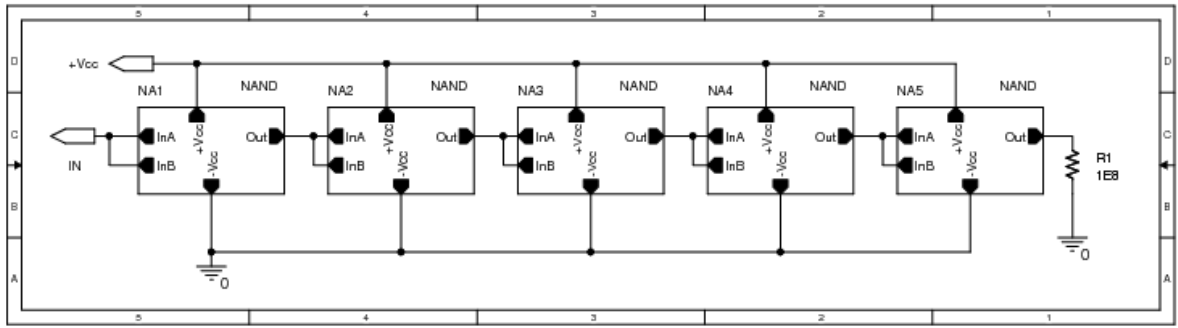

Figure 16. Schematic of the circuit used to simulate the behaviour of the NAND gate.

The result of the transient simulation in SPICE is shown in Fig. 17.

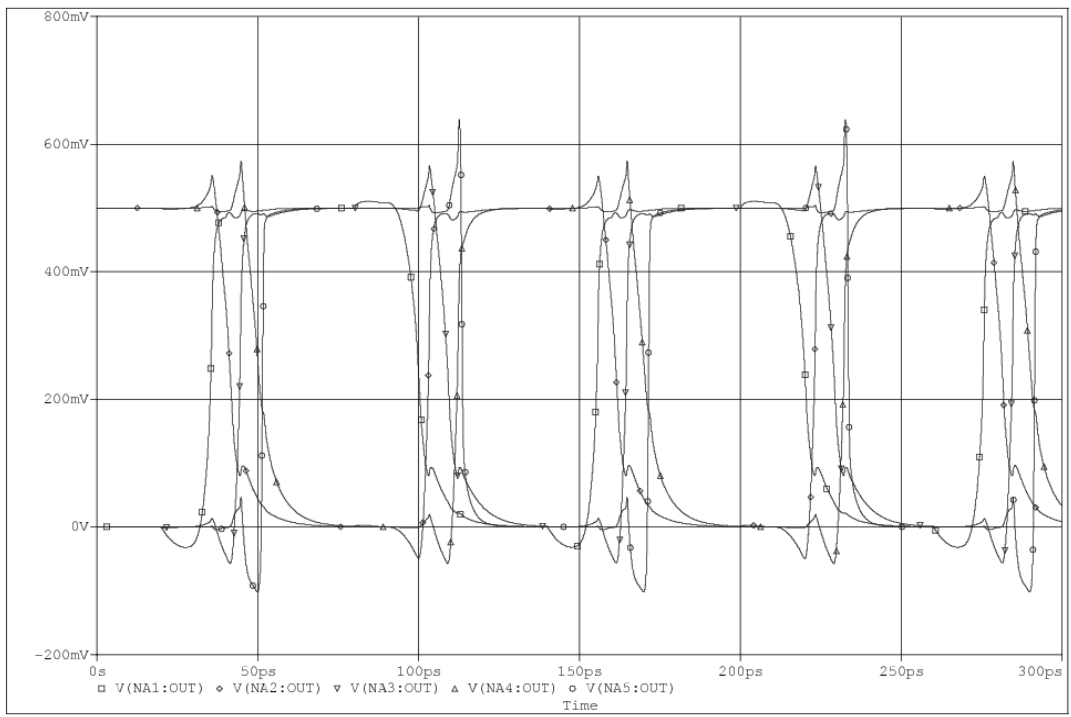

Figure 17. Transfer function of the circuit shown in Fig. 16 with SPICE, measured at the gate of first, second, third and fourth NAND, marked with circles, crosses, squares and triangles respectively.

In Verilog-A implementation we have obtained some differences and this comes from the incomplete implementation of intrinsic capacitance model in SPICE [7]. In fact, the number of sub-bands in our SPICE code has been considered equal to 1 to do not weigh down the software further, while in our Verilog-A implementation, the number of sub-bands can be defined as a parameter settable by the user (cfr. [6]).

# 6. CONCLUSIONS

Prediction through modelling forms the basis of engineering design. The computational power at the fingertips of the professional engineer is increasing enormously and techniques for computer simulation are changing rapidly. Engineers need models which relate to their design area and are adaptable to new design concepts. They also need efficient and friendly ways of presenting, viewing and transmitting the data associated with their models.

In this paper we have presented an exhaustive description **of** the basic types of CNTFETs. In particular we have reviewed two models, already proposed by us [3-5], which allow an easy implementation in circuit simulators, such as SPICE and Verilog-A, both in analog and in digital applications.

At last we have proposed and discussed two design examples: a CNTFET amplifier in CS configuration, for analog application, and NAND gates in a five stage chain, for digital one.

## Authors

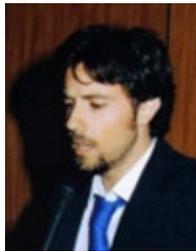

**Roberto Marani** received the Master of Science degree (*cum laude*) in Electronic Engineering in 2008 from Polytechnic University of Bari, where he received his Ph.D. degree in Electronic Engineering in 2012.

He worked in the Electronic Device Laboratory of Bari Polytechnic for the design, realization and testing of nanometrical electronic systems, quantum devices and FET on carbon nanotube. Moreover Dr. Marani worked in the field of design, modelling and experimental characterization of devices and systems for biomedical applications.

In December 2008 he received a research grant by Polytechnic University of Bari for his research activity. From February 2011 to October 2011 he went to Madrid, Spain, joining the Nanophotonics Group at Universidad Autónoma de Madrid, under the supervision of Prof. García-Vidal.

Currently he is involved in the development of novel numerical models to study the physical effects that occur in the interaction of electromagnetic waves with periodic nanostructures, both metal and dielectric. His research activities also include biosensing and photovoltaic applications.

Dr. Marani is a member of the COST Action MP0702 - Towards Functional Sub-Wavelength Photonic Structures, and is a member of the Consortium of University CNIT – Consorzio Nazionale Interuniversitario per le Telecomunicazioni.

Dr. Marani has published over 100 scientific papers.

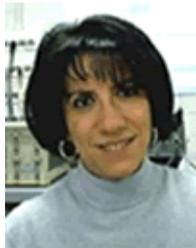

**Anna Gina Perri** received the Laurea degree *cum laude* in Electrical Engineering from the University of Bari in 1977. In the same year she joined the Electrical and Electronic Department, Polytechnic University of Bari, Italy, where she is Full Professor of Electronics from 2002.

From 2003 she has been associated with the National Institute of Nuclear Phisics (INFN) of Napoli (Italy), being a part of the TEGAF project: "Teorie Esotiche per Guidare ed Accelerare Fasci", dealing with the optimal design of resonance-accelerating cavities having very high potentials for cancer hadrontherapy.

In 2004 she was awarded the "Attestato di Merito" by ASSIPE (ASSociazione Italiana per la Progettazione Elettronica), Milano, BIAS'04, for her studies on electronic systems for domiciliary teleassistance.

Her current research activities are in the area of numerical modelling and performance simulation techniques of electronic devices for the design of GaAs Integrated Circuits and in the characterization and design of optoelectronic devices on PBG (Phothonic BandGap).

Moreover she works in the design, realization and testing of nanometrical electronic systems, quantum devices, FET on carbon nanotube and in the field of experimental characterization of electronic systems for biomedical applications.

Prof. Perri is the Head of the Electron Devices Laboratory of the Polytechnic University of Bari.

She has been listed in the following volumes: Who's Who in the World and Who's Who in Engineering, published by Marquis Publ. (U.S.A.).

She is author of over 250 journal articles, conference presentations, twelve books and currently serves as a Referee of a number of international journals.

Prof. Perri is the holder of two italian patents and the Editor of two international books.

She is also responsible for research projects, sponsored by the Italian Government.

Prof. Perri is a member of the Italian Circuits, Components and Electronic Technologies – Microelectronics Association, and an Associate Member of National University Consortium for Telecommunications (CNIT).